\documentclass[preprints,article,accept,moreauthors,pdftex]{Definitions/mdpi}
\pdfoutput=1

\firstpage{1}
\pubvolume{5}
\issuenum{6}
\articlenumber{142}
\pubyear{2019}
\copyrightyear{2019}
\history{Date: 3 June 2019}

\newcommand{\lc}[1]{\overset{\circ}{#1}\vphantom{#1}}

\usepackage{amsmath}
\usepackage{amsfonts}
\usepackage{amssymb}
\usepackage{ulem}

\Title{Flat connection for rotating vacuum spacetimes in extended teleparallel gravity theories}

\Author{Laur Järv $^{1}$, Manuel Hohmann $^{1}$, Martin Kr\v{s}\v{s}\'{a}k $^{1,2}$ and Christian Pfeifer $^{1}$}

\AuthorNames{Laur Järv, Manuel Hohmann, Martin Krššák, Christian Pfeifer}

\address{%
$^{1}$ \quad Laboratory of Theoretical Physics, Institute of Physics,
University of Tartu, W. Ostwaldi 1, 50411 Tartu, Estonia\\
$^{2}$ \quad Center for Gravitation and Cosmology, College of Physical Science and Technology, Yangzhou University, Yangzhou 225009, China}

\corres{Correspondence: laur.jarv@ut.ee}

\abstract{Teleparallel geometry utilizes Weitzenböck connection which has nontrivial torsion but no curvature and does not directly follow from the metric like Levi-Civita connection.
In extended teleparallel theories, for instance in $f(T)$ or scalar-torsion gravity, the connection must obey its antisymmetric field equations. So far only a few analytic solutions were known.
In this note we solve the $f(T,\phi)$ gravity antisymmetric vacuum field equations for a generic rotating tetrad ansatz in Weyl canonical coordinates, and find the corresponding spin connection coefficients. By a coordinate transformation we present the solution also in Boyer-Lindquist coordinates, often used to study rotating solutions in general relativity. The result hints for the existence of another branch of rotating solutions besides the Kerr family in extended teleparallel gravities.
}


\begin{document}

\section{Introduction}
The observation of gravitational waves engendered by the merger of two black holes \cite{Abbott:2016blz} and the very recent capture of the first image of a supermassive black hole \cite{Akiyama:2019cqa} underscore the importance of black hole research for the progress in fundamental physics \cite{Barack:2018yly}.
Determining the waveforms and shadows can allow to distinguish black holes form their more exotic dark compact cousins \cite{Cardoso:2019rvt,Bambi:2019tjh}, and probe the extensions and modifications of general relativity \cite{Berti:2015itd,Carballo-Rubio:2018jzw,Berti:2018vdi}.
The issue is whether the Kerr solution of a rotating black hole in general relativity (GR) can give a sufficiently good account of the observations, and how well can we test and rule out the possible alternatives. A systematic study to understand the properties of black holes and other compact dark objects in GR and beyond is therefore very much a call of the day.

In its geometric perspective GR assumes metric and torsion free Levi-Civita connection, and attributes the phenomenon of gravity to the curvature.
However, the choice of connection is a mathematical convention and not an independent property of spacetime. It is actually possible to rewrite the Einstein-Hilbert action of GR in terms of two different geometric concepts: the teleparallel (Weitzenböck) connection, which is endowed with nonzero torsion but vanishing curvature and vanishing nonmetricity, yielding teleparallel equivalent of general relativity (TEGR) where the dynamics of gravity can be expressed by the torsion \cite{AP,Maluf:2013gaa,Cai:2015emx} (see Refs.\ \cite{Hehl:1994ue,Blagojevic:2013xpa} for a general context), and the curvature free and torsion free symmetric teleparallel connection, yielding symmetric teleparallel equivalent of general relativity (STEGR) where the effects of gravity are encoded by nonmetricity \cite{Nester:1998mp,BeltranJimenez:2017tkd,Jarv:2018bgs}.
These three formulations of the theory (tentatively dubbed the ``geometrical trinity of gravity'' \cite{BeltranJimenez:2019tjy}) have equivalent field equations and hence possess the same black hole solutions, which may nevertheless offer insights into the discussions of e.g.\ gravitational energy, thermodynamics, etc. \cite{Maluf:2002zc,Lucas:2009nq,Maluf:2012na}. However, when in analogy to the curvature based $f(R)$ and scalar-tensor gravity, one extends the teleparallel theory by e.g.\ introducing an arbitrary function of the torsion scalar, $f(T)$ \cite{Bengochea:2008gz,Linder:2010py}, or nonminimally coupled scalar field in the action \cite{Geng:2011aj,Hohmann:2018rwf}, or generalizes the theory further \cite{Hohmann:2017duq,Hohmann:2018vle,Hohmann:2018dqh,Bahamonde:2017wwk,Bahamonde:2019gjk,Bahamonde:2019shr,Hohmann:2019gmt}, then the field equations start to differ from their curvature based counterparts, and consequently can offer new or modified solutions.

In contrast to the Levi-Civita connection, the teleparallel connection is not  entirely determined by the metric tensor. The constraints of vanishing curvature (``flatness'') and nonmetricity restrict the teleparallel connection coefficients in the most general case to depend on six functions, extra to the Levi-Civita part coming from the metric tensor.
How to appropriately interpret this additional freedom and determine teleparallel connection for a solution of interest, are among the central issues in the theory. In the case of TEGR, varying the action with respect to the independent part of the teleparallel connection gives an identically trivial result.
However, appropriately fixing teleparallel connection is still required for the correct definition of conserved charges \cite{Obukhov:2006sk,Lucas:2009nq,Krssak_Pereira2015} and physically relevant finite action \cite{Krssak_Pereira2015,Krssak:2015lba}. The issue is tied with the conceptual separation of gravitational and inertial effects, and the best proposal so far to find the teleparallel connection in TEGR proceeds by identifying the purely inertial connection pertaining to the limit where gravity is ``turned off'' \cite{Krssak:2015lba,Krssak:2017nlv,Krssak:2018ywd}.
In extended theories the variation of the action with respect to the independent part of the teleparallel connection yields six nontrivial antisymmetric equations \cite{Golovnev:2017dox,Krssak:2017nlv,Hohmann:2018rwf},  which is an essential ingredient in the covariant description of the theory \cite{Krssak:2015oua,Hohmann:2018rwf,Krssak:2018ywd}.  In the picture of local frames curvature can be made to vanish with zero spin connection, and the six functions parameterize the equivalence class of frames related to the frame with zero spin connection by local Lorentz transformations.

Speaking of connections, one also has to address the issue of how the matter fields couple to them. Including spinless matter as a scalar field or ideal fluid, as well as the electromagnetic field in a teleparallel theory is quite straightforward as they naturally only couple to the metric and feel the Levi-Civita connection. However, the prescription for the fermions coupling in TEGR has led to different approaches and conflicting opinions \cite{deAndrade:2001vx,Obukhov:2002tm,Mosna:2003rx,Maluf:2003fs,Mielke:2004gg,Obukhov:2004hv}. In the present work we focus upon vacuum field configurations in the exterior of a massive body or a black hole, hence we expect our results to hold in different possible scenarios in the matter sector.

There are several papers which consider rotating solutions in $f(T)$ teleparallel gravities. However, they proceed solving the field equations by accepting that the second derivative of $f$ with respect to $T$ vanishes \cite{Krssak_Pereira2015}, or adopting zero \cite{Bejarano:2014bca} or constant $T$ \cite{Nashed:2014iza,Nashed:2015pga}, which all render the field equations to those of TEGR. Therefore these results establish that the Kerr solution (as well as Schwarzschild \cite{Ferraro:2011ks,Tamanini:2012hg} and McVittie \cite{Bejarano:2017akj}) of GR survives as a universal solution in TEGR and in its generalizations.

It still leaves open an interesting question whether the family of rotating solutions is larger when we extend teleparallel gravity beyond TEGR. The aim of this short note is to take the first step and to solve the antisymmetric (connection) equations assuming a generic axially symmetric metric and scalar field in $f(T,\phi)$ extended teleparallel gravity. Indeed, we find a set of spin connection coefficients which imply nonconstant torsion scalar $T$, and at the same time get a constraint on the metric, which forces it to be different from Kerr. Obtaining the full solution would need solving the symmetric (metric) equations which depend on the particular form of the function $f$ and is beyond the scope of the present note. Meanwhile, the new axially symmetric connection we found is universal to the whole $f(T,\phi)$ family of theories.

In the next section \ref{sec:Covariant_formulation} we recollect the main formulas of teleparallel geometry and introduce $f(T,\phi)$ gravity in the formulation that is covariant under local Lorentz transformations. In Sec.\ \ref{sec:field_equations} we write down the field equations and discuss the possible approaches how to solve them with emphasis on treating the spin connection and the antisymmetric equations. Then in Sec.\ \ref{sec:Weyl_canonical} we start with an ansatz for a generic rotating spacetime and solve the antisymmetric field equations for the spin connection in Weyl canonical coordinates. The result will be presented in different Lorentz frames to illustrate the covariant formalism. The same solution is then transformed into Boyer-Lindquist coordinates in Sec.\ \ref{sec:Boyer_Lindquist}, and again exposed in different Lorentz frames.  Finally Sec.\ \ref{sec:Discussion} gives a summarizing discussion and outlook.


\section{Covariant formulation of $f(T,\phi)$ gravity}
\label{sec:Covariant_formulation}

The Weitzenböck connection ${\Gamma}^\rho{}_{\mu\nu}$ in teleparallel gravity is assumed to have vanishing curvature,
\begin{equation}\label{eqn:curvature tensor}
{R}^\rho{}_{\sigma \mu\nu} = \partial_{\mu}{\Gamma}^\rho{}_{\sigma\nu} - \partial_{\nu}{\Gamma}^\rho{}_{\sigma\mu} + {\Gamma}^\rho{}_{\lambda\mu}{\Gamma}^\lambda{}_{\sigma\nu} - {\Gamma}^\rho{}_{\lambda\nu}{\Gamma}^\lambda{}_{\sigma\mu} \equiv 0
\end{equation}
and vanishing nonmetricity,
\begin{equation} \label{eqn:nonmetricity}
{\nabla}_\rho g_{\mu\nu}=\partial_\rho g_{\mu\nu} - {\Gamma}^\lambda{}_{\mu\rho} g_{\lambda \nu} - {\Gamma}^\lambda{}_{\nu\rho} g_{\mu \lambda} \equiv 0 \,,
\end{equation}
but nonzero torsion,
\begin{equation}\label{eqn:torsion}
{T}^{\rho}{}_{\mu\nu} = {\Gamma}^{\rho}{}_{\nu\mu} - {\Gamma}^{\rho}{}_{\mu\nu}\,.
\end{equation}
The difference between the teleparallel connection and Levi-Civita connection whose coefficients are determined by the metric,
\begin{equation}
\lc{\Gamma}^{\rho}{}_{\mu\nu} = \frac{1}{2}g^{\rho\sigma}\left(\partial_{\mu}g_{\sigma\nu} + \partial_{\nu}g_{\mu\sigma} - \partial_{\sigma}g_{\mu\nu}\right)\,,
\end{equation}
is called the contortion tensor, which can be expressed as
\begin{equation}\label{eqn:contor}
K^{\rho}{}_{\mu\nu} = {\Gamma}^{\rho}{}_{\mu\nu} - \lc{\Gamma}^{\rho}{}_{\mu\nu} = \frac{1}{2}\left({T}_{\mu}{}^{\rho}{}_{\nu} + {T}_{\nu}{}^{\rho}{}_{\mu} - {T}^{\rho}{}_{\mu\nu}\right) \,.
\end{equation}
Note that the curvature $\lc{R}^\rho{}_{\sigma \mu\nu}$ of the Levi-Civita connection would still be nonvanishing in general, but the torsion $\lc{T}^{\rho}{}_{\mu\nu}$ is identically zero. (In some literature the teleparallel connection and the quantities computed from it are denoted by a filled overdot (e.g.\ \citep{AP,Krssak:2018ywd}) but we will omit this here. All quantities pertain to teleparallel connection unless marked otherwise.)

From the quantities above we may introduce the superpotential
\begin{equation}\label{eqn:suppot}
S_{\rho}{}^{\mu\nu} = K^{\mu\nu}{}_{\rho} - \delta_{\rho}^{\mu}T_{\sigma}{}^{\sigma\nu} + \delta_{\rho}^{\nu}T_{\sigma}{}^{\sigma\mu}\,
\end{equation}
and form a torsion scalar
\begin{equation}\label{eqn:torsscal}
T = \frac{1}{2}T^{\rho}{}_{\mu\nu}S_{\rho}{}^{\mu\nu}\,.
\end{equation}
The latter has a remarkable property
\begin{equation} \label{eqn:Ricci_theorem}
\lc{R}= -{T} + \frac{2}{\sqrt{g}}\partial_\mu(\sqrt{g}T_{\nu}{}^{\nu\mu}) \,.
\end{equation}
When submitted to a variational exercise here the last term becomes a boundary term that does not contribute to the field equations. This property allows to rewrite the Einstein-Hilbert action of GR in the teleparallel framework as
\begin{equation}\label{eqn:TEGRaction}
S_{TEGR} = \frac{1}{2\kappa^2}\int_M d^4x \, \sqrt{g} \, {T} \,.
\end{equation}
As the field equations derived from \eqref{eqn:TEGRaction} coincide with those of GR due to \eqref{eqn:Ricci_theorem}, the theory \eqref{eqn:TEGRaction} is known as teleparallel equivalent of general relativity.

While TEGR reproduces GR in an alternative geometrical setting, if we modify the action \eqref{eqn:TEGRaction} to a function of the torsion scalar or involve a nonminimally coupled field, we get a theory different from the curvature counterparts $f(R)$ or scalar-tensor gravity. Let us consider the action \cite{Hohmann:2018rwf}
\begin{equation}\label{eqn:gravaction}
S = \frac{1}{2\kappa^2}\int_M d^4x \, \sqrt{g} \left[f(T,\phi) + Z(\phi)g^{\mu\nu} \phi_{,\mu}\phi_{,\nu}\right]\,,
\end{equation}
which depends on two free functions \(f\) and \(Z\), while $2\kappa^2=16 \pi G_N$ sets the Newtonian gravitational constant. This action encompasses $f(T)$ gravity and teleparallel dark energy as particular cases. Here we assume that the connection in $T$ is teleparallel. An alternative would be to start with arbitrary connection and impose its flatness and metricity by adding suitable Lagrange multiplier terms in the action \cite{BeltranJimenez:2018vdo}. A complete physical theory is obtained by adding matter field actions to the gravitational actions \eqref{eqn:TEGRaction} or \eqref{eqn:gravaction}. In the literature on teleparallel gravity theories several inequivalent matter coupling models have been considered, as we mentioned in the introduction. Since we are interested in the solutions of the vacuum field equations we do not enter into this discussion here.

The conditions \eqref{eqn:curvature tensor} and \eqref{eqn:nonmetricity} reduce the freedom in teleparallel connection to six functions. This can be seen more elegantly in the tetrad formalism which allows to express quantities in anholonomic bases (frames) by relating the metric and tetrad $h^a{}_{\mu}$,
\begin{equation}\label{eqn:metric}
g_{\mu\nu} = \eta_{ab}h^a{}_{\mu}h^b{}_{\nu}\,,
\end{equation}
(where $\eta_{ab}=\mathrm{diag}(-1,1,1,1)$, \(h^a{}_{\mu}h_b{}^{\mu} = \delta^a_b\), \(h^a{}_{\mu}h_a{}^{\nu} = \delta_{\mu}^{\nu}\)) and introducing the spin connection ${\omega}^a{}_{b\mu}$
with
\begin{equation}\label{eqn:Gamma_omega}
{\Gamma}^{\rho}{}_{\mu\nu} = h_a{}^{\rho}\left(\partial_{\nu}h^a{}_{\mu} + {\omega}^a{}_{b\nu}h^b{}_{\mu}\right)\,.
\end{equation}
Tetrads can be used to transform between spacetime and frame indexes, e.g.
\begin{eqnarray}
T^a{}_{\mu\nu} &=& h^a{}_\rho T^\rho{}_{\mu\nu} =
\partial_\mu h^a_{\ \nu} -\partial_\nu h^a_{\ \mu}+\omega^a_{\ b\mu}h^b_{\ \nu}  -\omega^a_{\ b\nu}h^b_{\ \mu} \,,
\label{tordef}
\\
\label{eqn:curv}
{R}^a{}_{b\mu\nu} &=& h^a{}_\rho h^b{}_\sigma {R}^\rho{}_{\sigma\mu\nu} =
\partial_{\mu}{\omega}^a{}_{b\nu} - \partial_{\nu}{\omega}^a{}_{b\mu} + {\omega}^a{}_{c\mu}{\omega}^c{}_{b\nu} - {\omega}^a{}_{c\nu}{\omega}^c{}_{b\mu}\,.
\end{eqnarray}
It is important to realize that the relationship between spacetime and frame expressions is not unique,
since we are allowed to make local Lorentz transformations on the frame indices,
\begin{equation}
h'{}^a_{\ \mu}=\Lambda^a_{\ b}h^b_{\ \mu}, \quad \quad
\omega'{}^a_{\ b\mu}=\Lambda^a_{\ c}\omega^c_{\ d\mu}\Lambda_b^{\ d}+\Lambda^a_{\ c} \partial_\mu \Lambda_b^{\ c},\label{lortrans}
\end{equation}
(where $\Lambda_b{}^{d}$ is the inverse matrix of $\Lambda^{b}{}_d$) which leave the respective spacetime quantities intact.

On can easily translate the flatness condition \eqref{eqn:curvature tensor} into ${R}^a{}_{b \mu\nu}=0$ and solve the latter by asking the spin connection coefficients to vanish, ${\omega}^a{}_{b\mu}=0$. Then in all other frames obtained by a Lorentz transformation, the spin connection
\begin{equation}\label{eqn:Lambda_d_Lambda}
\omega{}^a{}_{b\mu}=\Lambda^a_{\ c} \partial_\mu \Lambda_b^{\ c}
\end{equation}
remains flat and metric.
This explains why the independent teleparallel connection is characterized by six functions, these are just the six independent Lorentz transformation parameters at every spacetime point. Note that vanishing spin connection does not imply zero spacetime connection because of \eqref{eqn:Gamma_omega}. If we had imposed ${\Gamma}^{\rho}{}_{\mu\nu}=0$, then both the curvature \eqref{eqn:curvature tensor} and torsion \eqref{eqn:torsion} would vanish. On the other hand vanishing spin connection and its Lorentz transformed versions imply the vanishing of curvature \eqref{eqn:curv}, but not of torsion \eqref{tordef}.

The specific frame where the spin connection vanishes defines the Weitzenböck gauge \cite{Obukhov:2002tm} (up to a global Lorentz transformation).
This gauge can be rather useful in calculations, but if one sets the spin connection to vanish as defining property of the theory, then local Lorentz covariance is lost and many problems ensue \cite{Li:2010cg, Sotiriou:2010mv}. If the spin connection is suppressed, the torsion tensor \eqref{tordef} and the quantities constructed from it will fail to transform covariantly under local Lorentz transformations. In particular, the actions \eqref{eqn:TEGRaction} an \eqref{eqn:gravaction} in such noncovariant formulation will remain invariant only under global Lorentz transformations whereby nonzero spin connection would otherwise not be generated by Eq.\ \eqref{lortrans}. In the covariant approach one allows nontrivial spin connection and considers all Lorentz frames on equal footing, the torsion \eqref{tordef} behaves correctly as a tensor and the action  \eqref{eqn:gravaction} is invariant under local Lorentz transformations \cite{Krssak:2015oua,Krssak:2018ywd}.

\section{Field equations}
\label{sec:field_equations}

The variation of the action \eqref{eqn:gravaction} with respect to the metric, or with respect to the tetrad and then symmetrizing the spacetime indices, yields the symmetric (scalar-vacuum) field equations \cite{Hohmann:2018rwf}
\begin{equation}\label{eqn:symfeq}
\frac{1}{2}fg_{\mu\nu} + \lc{\nabla}_{\rho}\left(f_TS_{(\mu\nu)}{}^{\rho}\right) - \frac{1}{2}f_TS_{(\mu}{}^{\rho\sigma}T_{\nu)\rho\sigma} - Z\phi_{,\mu}\phi_{,\nu} + \frac{1}{2}Zg_{\mu\nu}g^{\rho\sigma}\phi_{,\rho}\phi_{,\sigma} = 0 \,.
\end{equation}
Here $f_T$ denotes a derivative of $f(T)$ with respect to the torsion scalar $T$, etc., and comma a partial derivative with respect to the respective coordinate. One must keep in mind that
\begin{equation}
\label{eqn:connection condition f_T written out}
\partial_{\mu}f_T = f_{TT}\partial_{\mu}T + f_{T\phi}\partial_{\mu}\phi\,.
\end{equation}
The antisymmetric part of the tetrad field equations turns out to be equivalent to the equations coming from the variation of the action with respect to the teleparallel spin connection, in spacetime components \cite{Hohmann:2018rwf}
\begin{equation}\label{eqn:asymfeq}
\partial_{[\rho}f_TT^{\rho}{}_{\mu\nu]} = 0\,
\end{equation}
(here one needs to antisymmetrize the three lower indices and then sum over the repeating upper and lower index), or equivalently in terms of the frame fields \cite{Golovnev:2017dox,Krssak:2017nlv}
\begin{equation}
\label{eqn:concon}
\partial_{\mu}f_T\left[\partial_{\nu}\left(hh_{[a}{}^{\mu}h_{b]}{}^{\nu}\right) + 2hh_c{}^{[\mu}h_{[a}{}^{\nu]}{\omega}^c{}_{b]\nu}\right] = 0\,.
\end{equation}
Here $h=\det h^{a}{}_{\mu}$ is the volume factor equal to $\sqrt{g}$. There are six independent equations corresponding to the freedom in teleparallel connection, as encoded in the Lorentz matrices.
Finally the equation for the scalar field is
\begin{equation}\label{eqn:scalarfeq}
f_{\phi} - Z_{\phi}g^{\mu\nu}\phi_{,\mu}\phi_{,\nu} - 2Z\lc{\square}\phi = 0\,,
\end{equation}
where \(\lc{\square} = g^{\mu\nu}\lc{\nabla}_{\mu}\lc{\nabla}_{\nu}\) is the d'Alembert operator of the Levi-Civita connection.

It is quite obvious that if $f_{TT} \equiv 0$ we get a scalar field coupled to TEGR \eqref{eqn:TEGRaction}, if $f_{T\phi} \equiv 0$ we get a scalar field minimally coupled to $f(T)$ gravity, and if $f_{TT} \equiv 0$ as well as   $f_{T\phi} \equiv 0$ we get a scalar field minimally coupled to TEGR. Also, if $f_\phi \equiv 0$ the scalar field is massless and minimally coupled. Furthermore, one can show that when $T$ and $\phi$ are constant in spacetime and $f_\phi=0$, then the equations above also reduce to those of TEGR with a minimally coupled scalar field \cite{Hohmann:2018rwf}.
Notice, that the minimally coupled scalar field does not appear in the antisymmetric field equations \eqref{eqn:asymfeq}, and similarly would not any matter coupled to the metric only. On the contrary, if we had introduced extra matter directly coupled to the teleparallel connection (or torsion, which is equivalent), then the antisymmetric field equations would have acquired an additional source term (see e.g.\ \cite{BeltranJimenez:2018vdo}). However, we will not entertain the latter possibility here.

To complement the outline of the mathematical possibilities of how the antisymmetric field equation \eqref{eqn:asymfeq} can be satisfied, given in Ref.\ \cite{Hohmann:2018rwf}, let us briefly discuss here how to approach the field equations in practice.
Usually, wanting to study a specific physical system, one has some underlying symmetry in mind, which can be imposed on the dynamical fields of the theory to solve the field equations. In contrast to the curvature and Levi-Civita connection based theories, here this means that we look for tetrads and teleparallel (flat) connections, whereby both obey the symmetry demands. Combining the three requirements of symmetry, flatness and solving the equations, there are several appoaches one can pursue to determine the spin connection.

First, naively taking the most straightforward diagonal tetrad that corresponds to a metric with certain symmetry, and assuming vanishing spin connection, will generally not solve the antisymmetric field equations, except for a few really simple cases. A more fruitful method is to take that diagonal tetrad and apply a local Lorentz transformation \eqref{lortrans} on it, then still assuming vanishing spin connection go on to solve the antisymmetric field equations for the functions parametrizing the Lorentz transformation \cite{Tamanini:2012hg}. The solution will be a tetrad in the Weitzenböck gauge, while the other equivalent tetrad - spin connection pairs can be generated by arbitrary local Lorentz transformations. The spin connection in these pairs is guaranteed to be flat, since it obeys \eqref{eqn:Lambda_d_Lambda}, and also symmetric, since it is associated with a symmetric tetrad.

The second approach could be to take that diagonal tetrad and write the flat spin connection in terms of the Lorentz matrices as \eqref{eqn:Lambda_d_Lambda}, then solve the antisymmetric field equations for the functions parametrizing the Lorentz transformation. In essence, one is now solving the field equations assuming a non-Weitzenböck gauge. Like in the previous approach, the antisymmetric equations \eqref{eqn:asymfeq} would be second order differential equations for the six Lorentz functions, which could be rather complicated to tackle in a general case. But, like before, the situation will not be so bad, if one can guess that only one local Lorentz boost or rotation is needed. The resulting spin connection can be assessed for the symmetry properties afterwards.

The third approach would be to take the diagonal tetrad and try to solve the antisymmetric equations with the spin connection coefficients that satisfy the zero curvature condition \eqref{eqn:curv} simultaneously. The number of equations and different components one needs to solve for is large, but the equations contain now at most only first order derivatives of the spin connection coefficients. It might be helpful to guess that some spin connection coefficients should to be zero or assume not to depend on certain coordinates, for example by comparing with the spin connection in the ``zero gravity'' (Minkowski) limit where the torsion tensor vanishes. The resulting spin connection can be checked for the symmetry as a subsequent step.

Finally, the fourth approach would benefit from the recent Ref.\  \cite{Hohmann:2019nat}, where the mathematical details of the notion of symmetry in teleparallel geometry were worked out and several tetrads and teleparallel connections corresponding to various symmetries were derived. On can take a suitable tetrad - spin connection pair from Ref.\ \cite{Hohmann:2019nat}, whereby the symmetry and flatness are already implemented, and proceed to solve the antisymmetric field equations to fix the remaining freedom.

In the current work we have followed the third approach to solve the antisymmetric field equations, and used Ref.\  \cite{Hohmann:2019nat} to check the symmetry of the obtained spin connection. To complete the solution our result should be substituted into the symmetric and scalar field equations. However, solving these equations would require specifying the particular theory, i.e.\ the functions $f(T, \phi)$ and $Z(\phi)$, which is beyond the scope of the present note, and is left for future work.

\section{Rotating spacetime in Weyl canonical coordinates}
\label{sec:Weyl_canonical}

A general axially symmetric metric in Weyl canonical coordinates \((t, z, \rho, \varphi)\)  is given by \cite{Stephani:2003tm}
\begin{equation}
ds^2 = A^2 (dt-W d\varphi)^2-B^2(dz^2+d\rho^2)-\rho^2 A^{-2} d\varphi^2 \,,
\label{metric Weyl}
\end{equation}
where $A, B, W$ are arbitrary functions of $z$ and $\rho$. As one may easily verify, the metric \eqref{metric Weyl} can be obtained from the almost diagonal tetrad
\begin{equation}
h^{a}_{\ \mu} = \begin{pmatrix}
A & 0 & 0 & -AW \\
0 & -B & 0 & 0 \\
0 & 0 & -B & 0 \\
0 & 0 & 0 & -\rho A^{-1}
\end{pmatrix} \,.
\label{tetrad_bad_Weyl}
\end{equation}
via \eqref{eqn:metric}. To complement the setup, we assume the scalar field is also stationary and axially symmetric, $\phi(z,\rho)$.

We can determine the associated spin connection coefficients from the connection equation \eqref{eqn:concon} and the requirement of flatness, the vanishing of \eqref{eqn:curv}. The equation \eqref{eqn:concon} is given by a sum of two parts, one multiplied by $f_{TT}$ and another by $f_{T\phi}$, c.f.\ \eqref{eqn:connection condition f_T written out}. We proceed by finding a solution that would satisfy both these parts separately, so that the result would remain independent of the function $f$ that specifies a particular theory. It is a lengthy computation, but assuming that the spin connection components are at most functions of $z,\rho$, and demanding that the coefficients of $\partial^2_z A$, $\partial^2_\rho A$, $\partial^2_z B$, $\partial^2_\rho B$, $ \partial^2_z W$, $\partial^2_\rho W$ vanish independently (since $A, B$ and $W$ are arbitrary), yields a solution with only two nonzero components
\begin{equation}
\omega^{\hat{2}}_{\ \hat{3} \varphi} = - \omega^{\hat{3}}_{\ \hat{2} \varphi} = -1 \,.
\label{connection_Weyl}
\end{equation}
This spin connection solves five of the six equations \eqref{eqn:curv}, while the remaining one can be satisfied by
\begin{equation}
B = \frac{1}{A} \,.
\label{AB}
\end{equation}
Comparing with Ref.\  \cite{Hohmann:2019nat} we may confirm that the spin connection \eqref{connection_Weyl} is a particular case of a generic axially symmetric connection.

In fact, we could have obtained the spin connection \eqref{connection_Weyl} by the procedure originally suggested for TEGR, namely ``turning off'' gravity so that only purely inertial connection remains \cite{Krssak:2015lba,Krssak:2017nlv,Krssak:2018ywd}. Indeed, demanding that the torsion tensor components vanish for a reference tetrad with $A=B=1,W=0$ which corresponds to Minkowski space in cylindrical coordinates, immediately gives \eqref{connection_Weyl}.
Physically this spin connection is easy to interpret as accounting for the inertial effects stemming from the cylindrical coordinate system. However in extended teleparallel theories the spin connection must also satisfy the equations \eqref{eqn:concon}, and this brings in an extra demand \eqref{AB}. The latter is rather difficult to interpret at this point, especially since  imposing \eqref{AB} in \eqref{metric Weyl} precludes Kerr solution (c.f.\ for instance \cite{Jones:2005hj}).

It is possible to find a local Lorentz transformation that makes the spin connection to vanish. In other words we need to determine $\Lambda^a_{\ b}$ that generates \eqref{connection_Weyl} via \eqref{eqn:Lambda_d_Lambda}. It turns out to be a simple rotation
\begin{equation}
\Lambda^a_{\ b} = \begin{pmatrix}
1 & 0 & 0 & 0 \\
0 & 1 & 0 & 0 \\
0 & 0 & \cos \varphi & \sin \varphi \\
0 & 0 & -\sin \varphi & \cos \varphi
\end{pmatrix}
\,.
\label{rotation_Weyl}
\end{equation}
Applying the inverse rotation
\begin{equation}
(\Lambda^{-1})^a_{\ b} = \begin{pmatrix}
1 & 0 & 0 & 0 \\
0 & 1 & 0 & 0 \\
0 & 0 & \cos \varphi & -\sin \varphi \\
0 & 0 & \sin \varphi & \cos \varphi
\end{pmatrix}
\end{equation}
in \eqref{lortrans} makes the spin connection to vanish and transforms the tetrad \eqref{tetrad_bad_Weyl} into
\begin{equation}
h^{a}_{\ \mu} = \begin{pmatrix}
A & 0 & 0 & -AW \\
0 & -\frac{1}{A} & 0 & 0 \\
0 & 0 & -\frac{\cos \varphi}{A} & \frac{\rho \sin \varphi}{A} \\
0 & 0 & -\frac{\sin \varphi}{A} & -\frac{\rho \cos \varphi}{A}
\end{pmatrix} \,.
\label{tetrad_good_Weyl}
\end{equation}
Indeed, the tetrad \eqref{tetrad_good_Weyl} satisfies the condition \eqref{eqn:concon} with vanishing spin connection. In the parlance of the older noncovariant ``pure tetrad'' approach to teleparallel gravity the nondiagonal tetrad \eqref{tetrad_good_Weyl} would be called a ``good tetrad'' in analogy with the nondiagonal tetrads for the spherically symmetric spacetimes \cite{Ferraro:2011ks,Tamanini:2012hg}.
In the newer covariant approach the tetrad \eqref{tetrad_good_Weyl} is called a ``proper tetrad'' and interpreted to correspond to a frame where the inertial effects are not present. The spherically symmetric analogue for the diagonal tetrad \eqref{tetrad_bad_Weyl} and nontrivial spin connection \eqref{connection_Weyl} was given in Ref.\ \cite{Krssak:2015oua}. In the spherically symmetric case an extra condition like \eqref{AB} did not arise.

As a remark, let us note that if we took the ``good'' tetrad solution \eqref{tetrad_good_Weyl} and made a Lorentz transform \eqref{rotation_Weyl} we would have gotten the tetrad \eqref{tetrad_bad_Weyl}. However, the latter without spin connection would not solve the field equations. In the old noncovariant approach which assumed identically zero spin connection, this phenomenon was a source of some puzzlement and led to the terminology whereby the tetrad \eqref{tetrad_bad_Weyl} would be called ``bad'' \cite{Tamanini:2012hg}.
In the covariant understanding of the theory the tetrad \eqref{tetrad_bad_Weyl} is in no sense bad or unphysical, it just solves the field equations in combination with a nonzero spin connection \eqref{connection_Weyl} engendered by the same Lorentz transformation \eqref{rotation_Weyl} via \eqref{eqn:Lambda_d_Lambda}.

We can also record that the torsion scalar \eqref{eqn:torsscal} corresponding to the diagonal tetrad \eqref{tetrad_bad_Weyl} and spin connection \eqref{connection_Weyl}, or equivalently to the nondiagonal tetrad \eqref{tetrad_good_Weyl} and vanishing spin connection (all with \eqref{AB} implied) is given by
\begin{equation}
T = \frac{A^6 \left( (\partial_z W)^2 + (\partial_\rho W)^2 \right) - 4 \rho^2 \left( (\partial_z A)^2 + (\partial_\rho A)^2 \right) }{2 \rho^2} \,.
\end{equation}
The torsion scalar is not constant in general, hence does not render the equations automatically to those of TEGR. Therefore there is a possibility to obtain new solutions that are not present in TEGR (and correspondingly, in GR). The functions $A, W$ still need to be determined by solving the symmetric field equations \eqref{eqn:symfeq} and the scalar field equation \eqref{eqn:scalarfeq}, which cannot be undertaken without specifying the theory (i.e.\ the function $f$).

\section{Rotating spacetime in Boyer-Lindquist coordinates}
\label{sec:Boyer_Lindquist}

To facilitate comparisons with the most common presentation of the Kerr solution, let us convert the solution from Weyl canonical coordinates into the Boyer-Lindquist coordinates \((t, r, \vartheta, \varphi)\) by \cite{Stephani:2003tm}
\begin{equation}
t=t \,, \qquad  \rho= \sqrt{\Delta} \sin \vartheta \,, \qquad z=(r-m) \cos \theta \,, \qquad \varphi=\varphi
\label{Weyl_Boyer_Lindquist}
\end{equation}
where
\begin{equation}
\Delta = {r^2-2mr+a^2} \,.
\end{equation}
Here we should stress that while in the Kerr solution the parameters $m$ and $a$ carry a clear physical meaning, at this point here they are just some constants in the coordinate transformation, without a physical interpretation yet.
We can make the coordinate transformation \eqref{Weyl_Boyer_Lindquist} to the nondiagonal tetrad \eqref{tetrad_good_Weyl} to obtain
\begin{equation}
h^{a}_{\ \mu} = \begin{pmatrix}
A & 0 & 0 & -AW \\
0 & -\frac{\cos \vartheta}{A} & \frac{(r-m)\sin \vartheta}{A} & 0 \\
0 & -\frac{(r-m) \cos \varphi \, \sin \vartheta}{A \sqrt{\Delta}} & -\frac{\sqrt{\Delta} \cos \varphi \, \cos \vartheta }{A} & \frac{\sqrt{\Delta} \sin\varphi \, \sin \vartheta }{A} \\
0 & -\frac{(r-m)\sin \varphi \, \sin \vartheta}{A \sqrt{\Delta}} & -\frac{\sqrt{\Delta} \sin \varphi \cos \vartheta}{A} & -\frac{\sqrt{\Delta} \cos \varphi \, \sin \vartheta}{A}
\end{pmatrix} \,,
\label{tetrad_good_BL}
\end{equation}
where $A, W$ are now functions of $r, \vartheta$.
It can be checked that the transformed nondiagonal tetrad \eqref{tetrad_good_BL} still satisfies \eqref{eqn:concon} with vanishing spin connection, as it should. The torsion scalar is now given by
\begin{equation}
T = \frac{A^6 \left( \frac{(\partial_r W)^2}{\sin^2 \vartheta} + \frac{(\partial_\vartheta W)^2}{\Delta \sin^2\vartheta} \right) - 4  \left( \Delta (\partial_r A)^2 + (\partial_\vartheta A)^2 \right) }{2 \Omega} \,,
\end{equation}
where
\begin{equation}
\Omega = r^2-2mr+m^2+(a^2-m^2)\cos^2 \vartheta \,.
\end{equation}

We may seek to find a Lorentz frame where the tetrad \eqref{tetrad_good_BL} diagonalizes as much as possible. For that let us
rotate it by making a Lorentz transformation with \eqref{rotation_Weyl}.
The result is
\begin{equation}
h^{a}_{\ \mu} = \begin{pmatrix}
A & 0 & 0 & -AW \\
0 & -\frac{\cos \vartheta}{A} & \frac{(r-m)\sin \vartheta}{A} & 0 \\
0 & -\frac{(r-m) \sin \vartheta}{A \sqrt{\Delta}} & -\frac{\sqrt{\Delta} \cos \vartheta }{A} & 0 \\
0 & 0 & 0 & -\frac{\sqrt{\Delta} \sin \vartheta}{A}
\end{pmatrix} \,.
\label{tetrad_bad_BL}
\end{equation}
The latter tetrad is obviously associated with the spin connection \eqref{connection_Weyl}. Indeed, it can be checked that together the tetrad \eqref{tetrad_bad_BL} and the spin connection \eqref{connection_Weyl} satisfy the connection equation \eqref{eqn:concon}.
Another rotation
\begin{equation}
\Lambda^a_{\ b} = \begin{pmatrix}
1 & 0 & 0 & 0 \\
0 & \frac{\sqrt{\Delta} \cos\vartheta}{\sqrt{\Omega}} & \frac{(r-m) \sin\vartheta}{\sqrt{\Omega}} & 0 \\
0 & - \frac{(r-m) \sin\vartheta}{\sqrt{\Omega}} & \frac{\sqrt{\Delta} \cos\vartheta}{\sqrt{\Omega}} & 0 \\
0 & 0 & 0 & 1
\end{pmatrix}
\label{rotation_to_diagonal_BL}
\end{equation}
takes the rotated tetrad \eqref{tetrad_bad_BL} into a ``diagonal'' form
\begin{equation}
h^{a}_{\ \mu} = \begin{pmatrix}
A & 0 & 0 & -AW \\
0 & -\frac{\sqrt{\Omega}}{A\sqrt{\Delta}} & 0 & 0 \\
0 & 0 & -\frac{\sqrt{\Omega}}{A} & 0 \\
0 & 0 & 0 & -\frac{\sqrt{\Delta} \sin\vartheta}{A}
\end{pmatrix} \,.
\label{tetrad_diagonal_BL}
\end{equation}
This last ``diagonal'' tetrad \eqref{tetrad_diagonal_BL} is associated with the spin connection
\begin{gather}
\omega^{\hat{1}}_{\ \hat{2}r} = - \omega^{\hat{2}}_{\ \hat{1}r} = \frac{(m^2 - a^2)\sin\vartheta\cos\vartheta}{\Omega\sqrt{\Delta}} \,, \qquad
\omega^{\hat{1}}_{\ \hat{2}\vartheta} = - \omega^{\hat{2}}_{\ \hat{1}\vartheta} = - \frac{(r-m) \sqrt{\Delta}}{\Omega} \,,
\nonumber \\
\omega^{\hat{1}}_{\ \hat{3}\varphi} = - \omega^{\hat{3}}_{\ \hat{1}\varphi} = - \frac{(r-m) \sin\vartheta}{\sqrt{\Omega}} \,, \qquad
\omega^{\hat{2}}_{\ \hat{3}\varphi} = - \omega^{\hat{3}}_{\ \hat{2}\varphi} = - \frac{\sqrt{\Delta} \cos\vartheta}{\sqrt{\Omega}}
\label{connection_diagonal_BL}
\end{gather}
which can be generated by \eqref{eqn:Lambda_d_Lambda} where the Lorentz transformation is a composition of \eqref{rotation_Weyl} and \eqref{rotation_to_diagonal_BL}, or by Lorentz transforming \eqref{lortrans} the connection \eqref{connection_Weyl} by \eqref{rotation_to_diagonal_BL}. The tetrad \eqref{tetrad_diagonal_BL} together with the spin connection \eqref{connection_diagonal_BL} again satisfies the connection equation \eqref{eqn:concon}.

To be sure, all the tetrads \eqref{tetrad_good_BL}, \eqref{tetrad_bad_BL}, and \eqref{tetrad_diagonal_BL} generate a rotating metric
\begin{equation}
\label{eqn:metric_in_Boyer_Lindquist}
ds^2 = A^2 (dt - W d\varphi)^2 - \frac{\Omega}{A^2} \left(\frac{1}{\Delta} dr^2 - d\vartheta^2 \right) - \frac{\Delta}{A^2} \sin^2\vartheta \, d\varphi^2 \,.
\end{equation}
Here the two arbitrary functions $A,W$ need to be determined by the tetrad field equations \eqref{eqn:symfeq} and the scalar field equation \eqref{eqn:scalarfeq}. By inspecting the $tt$ and $rr$ components, it becomes clear that a metric in the form of \eqref{eqn:metric_in_Boyer_Lindquist} can not be exactly congruent to the Kerr metric. The origin for this feature is the condition \eqref{AB}. We were seeking solutions different from TEGR and on purpose tried to solve the connection equation \eqref{eqn:concon} by not imposing the torsion scalar $T$ to be zero or constant. The Kerr solution is still a solution in $f(T,\phi)$ gravity, but it can be found on a different branch of connections, namely those which are common with TEGR \cite{Bejarano:2014bca}.

Finally let us remark that the spacetime components of the teleparallel connection can be computed via Eq.\ \eqref{eqn:Gamma_omega} from the tetrad - spin connection pairs given above. Although the tetrad and spin connection change from one Lorentz frame to another, the spacetime components of the connection remain unaffected, just like the metric. Although we are not going to write them out explicitly, the spacetime components of the teleparallel connection are not arbitrary. We have  determined them by solving the antisymmetric field equations, while the exact form of the free functions should get fixed by the symmetric field equations and the scalar field equation.

\section{Discussion}
\label{sec:Discussion}

Solving extended teleparallel gravity means not just finding the metric but also the independent teleparallel connection, thus satisfying both the symmetric and antisymmetric field equations. In contrast to GR and arguably also TEGR, a solution in $f(T)$ or scalar-torsion gravity is incomplete without determining the connection. When we work in the formalism of local frames, a solution would entail fixing a tetrad and the associated spin connection. The covariant formulation here fluently accommodates an equivalence class of different Lorentz frames by allowing nontrivial spin connection.

So far only a few explicit examples of teleparallel connections were known, which solve the antisymmetric field equations independent of the function $f(T,\phi)$, viz.\ the spherically symmetric \cite{Tamanini:2012hg,Krssak:2015oua} and cosmological spacetimes \cite{Tamanini:2012hg,Krssak:2015oua, Hohmann:2018rwf,Capozziello:2018hly, Hohmann:2019nat}. In this paper we presented an example of a teleparallel connection that solves the antisymmetric field equations for rotating spacetimes. It imposes an additional constraint on the metric, which renders the geometry different from the Kerr solution. Obtaining a full solution would also need tackling the symmetric and scalar field equations, but our result could be a first indication that $f(T,\phi)$ theories harbor an alternative branch of rotating black hole solutions besides the Kerr spacetime, which has been shown to be universal in TEGR and beyond \cite{Bejarano:2014bca}.
Since the antisymmetric field equations are not altered by the presence of minimally coupled matter, our result will still hold when such matter is included.

We presented the solution in Weyl canonical and Boyer-Lindquist coordinates, and expressed it in different Lorentz frames, explaining how the Lorentz covariant formulation of the theory works. The presented results may be interesting in a further analysis of the tetrads and spin connections, especially in trying to understand better how the inertial effects manifest for different observers (building up upon e.g.\ Ref.\ \cite{Maluf:2007qq}).
Enlarging the repertoire of analytic solutions will hopefully also inform and complement the discussion on more fundamental questions, like the nature of the (extra) degrees of freedom in the theory \cite{Ferraro:2018tpu,Ferraro:2018axk,Golovnev:2018wbh,Blixt:2019mkt} and the uniqueness of the connection for a given metric (compare the open universe solutions in Refs.\ \cite{Capozziello:2018hly} and \cite{Hohmann:2018rwf,Hohmann:2019nat}).
Once a full solution is obtained, it will be certainly interesting to analyse it in comparison with Kerr and its cousins in extended gravity theories (starting from e.g.\ Refs.\ \cite{Mccrea:1987rr,Bakler:1988nq}).
Hopefully the present work will also contribute to the prospective use of the phenomenology of black holes to test teleparallel theories, like cosmology \cite{Hohmann:2017jao,Capozziello:2017uam,Nunes:2018xbm,Abedi:2018lkr,Basilakos:2018arq,DAgostino:2018ngy,El-Zant:2018bsc,Capozziello:2019cav} and gravitational waves \cite{Abedi:2017jqx,Farrugia:2018gyz, Hohmann:2018jso, Cai:2018rzd, Nunes:2018evm}.

\vspace{6pt}

\authorcontributions{Conceptualization, validation, formal analysis, writing--review and editing: L.J., M.H., M.K, C.P.; funding acquisition: L.J., M.H.; methodology, investigation, writing--original draft preparation, supervision,  project administration: L.J.}

\funding{This research was funded by the Estonian Research Council through the projects IUT02-27, PUT790, and PRG356, as well as by the European Regional Development Fund through the Center of Excellence TK133 “The Dark Side of the Universe”.}

\conflictsofinterest{The authors declare no conflict of interest.}

\abbreviations{The following abbreviations are used in this manuscript:\\

\noindent
\begin{tabular}{@{}ll}
GR & General relativity\\
STEGR & Symmetric teleparallel equivalent of general relativity\\
TEGR & Teleparallel equivalent of general relativity\\
\end{tabular}}

\reftitle{References}


\end{document}